\documentclass[sigconf,10pt]{acmart}
\usepackage[english]{babel}
\usepackage{blindtext}
\renewcommand\footnotetextcopyrightpermission[1]{} 
\usepackage{siunitx}            
\usepackage{comment}
\usepackage{courier}            
\usepackage[scaled]{helvet}     
\usepackage{url}                
\usepackage{enumitem}           
\usepackage{xspace}
\usepackage{lipsum}
\usepackage{graphicx}
\usepackage{cleveref}

\definecolor{bg}{rgb}{0.5,0.5,0.5}
\definecolor{bggray}{rgb}{0.9, 0.9, 0.9}
\definecolor{brickred}{rgb}{0.8, 0.25, 0.33}
\definecolor{brightpink}{rgb}{1.0, 0.0, 0.5}
\definecolor{bluegray}{rgb}{0.4, 0.6, 0.8}

\newcommand{\sys}{\textsc{Wasm-bpf}\xspace}

\usepackage{amssymb} 
\usepackage{pifont}  
  {\begin{enumerate}[itemsep=-0pt, parsep=0pt, topsep=0pt, leftmargin=1pc]}
  {\end{enumerate}}

\newenvironment{smitemize}%
  {\begin{list}{$\bullet$}%
    {\setlength{\parsep}{0pt}%
      \setlength{\topsep}{0pt}%
      \setlength{\leftmargin}{2pc}%
      \setlength{\itemsep}{1pt}}}
  {\end{list}}
\newlength{\mintednumbersep}
\AtBeginDocument{%
  \sbox0{\tiny00}%
  \setlength\mintednumbersep{\parindent}%
  \addtolength\mintednumbersep{-\wd0}%
}

\begin{document}

\title{\sys: Streamlining eBPF Deployment in Cloud Environments with WebAssembly}

%
%

\author{Yusheng Zheng}
\affiliation{%
  \institution{eunomia-bpf Community}
  \country{China}
}
\email{yunwei356@gmail.com}

\author{Tong Yu}
\affiliation{%
  \institution{eunomia-bpf Community}
  \country{China}
}
\email{yunwei356@gmail.com}

\author{Yiwei Yang}
\affiliation{%
  \institution{UC Santa Cruz}
  \country{China}
}
\email{yyang363@ucsc.edu}

\author{Andrew Quinn}
\affiliation{%
  \institution{UC Santa Cruz}
  \country{USA}
}
\email{aquinn@ucsc.edu}


\begin{abstract}

The extended Berkeley Packet Filter (eBPF) is extensively utilized for observability and performance analysis in cloud-native environments. However, deploying eBPF programs across a heterogeneous cloud environment presents challenges, including compatibility issues across different kernel versions, operating systems, runtimes, and architectures. Traditional deployment methods, such as standalone containers or tightly integrated core applications, are cumbersome and inefficient, particularly when dynamic plugin management is required. To address these challenges, we introduce Universal BPF (\sys), a lightweight runtime on WebAssembly (Wasm) and the WebAssembly System Interface (WASI). Leveraging Wasm's platform independence and WASI's standardized system interface, with enhanced relocation for different architectures, \sys ensures cross-platform compatibility for eBPF programs. It simplifies deployment by integrating with container toolchains, allowing eBPF programs to be packaged as Wasm modules that can be easily managed within cloud environments. Additionally, \sys supports dynamic plugin management in WebAssembly. Our implementation and evaluation demonstrate that \sys introduces minimal overhead compared to native eBPF implementations while simplifying the deployment process.

\end{abstract}

\maketitle

\section{Introduction}\label{sec:introduction}

The extended Berkeley Packet Filter (eBPF) has emerged as a powerful technology for observability and performance in cloud-native environments\cite{sharmaebpf,he2023cross,shen2023network,zhou2023electrode,zhou2024dint,Zhong22}. By enabling the execution of custom bytecode within the kernel, eBPF provides a versatile mechanism for monitoring and modifying system behavior without requiring kernel modifications. This capability is particularly valuable in containerized and cloud ecosystems where dynamism and scalability are critical.

However, deploying eBPF programs heterogeneously at scale presents several significant challenges. eBPF runtimes extend beyond the Linux kernel, encompassing environments such as Windows\cite{eBPFWindows} and FreeBSD kernels, as well as userspace eBPF runtimes like ubpf\cite{ubpf}, rbpf\cite{rbpf}, and bpftime\cite{zheng2023bpftime}. Additionally, eBPF must be compatible across multiple operating system versions, each with unique kernel data structures, eBPF features, and capabilities, resulting in inconsistent application behavior. Although Compile Once - Run Everywhere (CO-RE) support can mitigate some issues, it also introduces further complexity to the deployment process.

Current methods for deploying eBPF programs in containers are often inefficient and cumbersome. Standalone containers, commonly used for eBPF deployment, do not optimize for the typically small and namespace-unaware nature of eBPF programs, leading to resource inefficiencies and management difficulties. Large-scale projects like Cilium\cite{cilium}, Pixie\cite{pixie}, Tetragon\cite{tetragon}, and Deepflow\cite{shen2023network} often integrate monitoring or management tools directly within the core application or "control plane", which can complicate management, especially in dynamic environments requiring frequent updates. Comprehensive observability agents like Deepflow necessitate dynamic plugin management to adapt to varying scenarios, a feature not adequately supported by traditional tools. Alternatively, some approaches use Remote Procedure Calls (RPCs) to interface between control plane applications and dedicated BPF daemons, as seen in tools like bpfman\cite{bpfman}, Inspektor-Gadget\cite{inspektor}, and bpfd\cite{bpfd}. While effective for specific eBPF tools, these methods are better suited for smaller-scale deployments and do not fully address broader compatibility and scalability issues.

To illustrate, consider a scenario where we need to monitor network traffic across a heterogeneous cloud environment with nodes running different kernels and instruction sets. These nodes might include older versions of Linux that do not natively support eBPF, newer versions of Linux with full eBPF support, Windows, and servers with diverse architectures such as ARM64 and x86. To ensuring that the observed data is aware of container metadata and correlates with other data sources, such as API requests, developers may integrate this monitoring as a plugin with observability agents like Deepflow allows for comprehensive data collection and correlation. With traditional methods, deploying eBPF programs in this diverse environment would require significant effort to ensure compatibility across different kernels and architectures. This process could involve building and maintaining separate versions of eBPF programs for each environment, leading to increased complexity and potential errors.

To address these challenges, we introduce \sys, a lightweight runtime based on WebAssembly (Wasm)~\cite{ramesh2023stop,webassembly,Haas17} and the WebAssembly System Interface (WASI)~\cite{wasi}. \sys is designed to streamline the distribution and compatibility of eBPF programs in cloud-native environments. By leveraging Wasm's platform independence and designing a standardized system interface based on WASI, \sys ensures seamless cross-platform compatibility for both eBPF and control-plane programs. It includes WebAssembly libraries for C, Go, and Rust toolchains, and provides runtime support for loading and executing eBPF programs from the Wasm module. \sys features an automatic mechanism for selecting appropriate runtimes—whether in the kernel or userspace—and supports various eBPF features such as ring buffers and perf events. Additionally, \sys enhances Compile Once - Run Everywhere (CO-RE) capabilities, ensuring support across different architectures and runtimes, without serialization overhead for complex data types shared between user-space Wasm runtime and kernel eBPF runtime. Moreover, \sys integrates seamlessly with container toolchains, simplifying the deployment and orchestration of eBPF programs within cloud environments. This integration addresses the complexities associated with current deployment methods, offering a more efficient and scalable solution.

In summary, our contributions are:

\begin{smitemize}
\item \textbf{Heterogeneous Platform Compatibility}: \sys introduces a lightweight runtime ensuring compatibility across different architectures and operating systems for both eBPF and control-plane programs, based on standard WASI and Wasm. It includes features like arch-aware relocation, automatic BTF preparing and facilitates selecting between different runtimes and eBPF features. It can run various eBPF programs with binary compatible on different OS and runtimes.
\item \textbf{Minor Performance Overhead} We analyze the additional overhead introduced by the compatibility layer, including costs associated with BPF syscalls, ring buffer handling, and initialization latency.
\item \textbf{Faster and Easier Deployment} \sys integrates seamlessly with container toolchains, simplifying the deployment and orchestration of eBPF programs in cloud-native environments.
\end{smitemize}
\section{Background}\label{sec:background}

This section overviews eBPF and WebAssembly (Wasm) applications and their importance in cloud computing environments.

\subsection{eBPF application}\label{sec:ebpf-core}

Initially developed for the Linux kernel to execute custom bytecode, eBPF has expanded to userspace runtimes like ubpf\cite{ubpf}, rbpf\cite{rbpf}, and bpftime\cite{zheng2023bpftime}, as well as in other operating system kernels such as Windows\cite{windows-ebpf} and FreeBSD\cite{freebsd}. An eBPF application typically consists of kernel bytecode and a userspace control plane application that loads this bytecode into the kernel and communicates through eBPF maps. The userspace application loads the eBPF bytecode, attaches it to the tracepoint sched\_wakeup, creates BPF maps for communication, and periodically reads these maps to print the histogram. This structure allows dynamic and flexible interaction between userspace applications and eBPF programs running within the kernel.



\subsection{Wasm and WASI}\label{sec:wasm-wasi}

WebAssembly (Wasm)\cite{webassembly} is a binary instruction format designed for safe and efficient execution of code across different environments. Originally developed for web browsers, Wasm has evolved to support a wide range of applications beyond the browser, including server-side\cite{faasm,faabric}, embedded systems\cite{pereira2021arena} and containers\cite{makitalo2021webassembly}. Its key advantages include platform independence, security through sandboxing, and near-native performance. By compiling programs into a portable binary format, Wasm ensures that they can run consistently across diverse architectures and operating systems without requiring source code modifications. This makes Wasm an ideal foundation for building cross-platform compatibility layers, such as \sys.

The WebAssembly System Interface (WASI) extends the capabilities of Wasm by providing a standardized API for interacting with the underlying operating system. WASI abstracts system-level functionalities such as file and network I/O, enabling Wasm modules to perform operations that would otherwise require platform-specific code. This abstraction layer allows developers to write code that is both portable and powerful, leveraging the full capabilities of the host environment while maintaining cross-platform compatibility.

\section{eBPF Deployment Challenges}\label{sec:challenges}

This section outlines key challenges in deploying eBPF programs, including compatibility issues, cloud deployment complexities, and versioning difficulties.

\subsection{Compatibility issues Across Different Runtimes and Architectures}

eBPF was initially designed for the Linux kernel, but its utility has led to implementations in other environments, such as Windows\cite{windows-ebpf} and FreeBSD kernels\cite{freebsd}, as well as userspace runtimes. Despite these diverse and heterogeneous runtimes adhering to the same instruction standard, they lack a unified standard for loading and deploying eBPF programs. Each operating system and runtime employs different methods to load eBPF bytecode. For instance, Linux utilizes bpf-related syscalls that directly accept eBPF bytecode. The eBPF-for-Windows project provides source code level compatibility with the Linux kernel and a libbpf API, but still requires some code modifications to most applications. Userspace runtimes like uBPF and rBPF do not support using libbpf as a loader, making them incompatible with existing control plane applications. Different runtimes also support different features. For example, uprobe can be used by userspace eBPF runtimes and the Linux kernel but is not available on Windows. Additionally, varying support for eBPF features like ring buffers or perf events across different Linux versions complicates deployment. The lack of a standardized method for recognizing and distributing these features further complicates deployment. The Compile Once - Run Everywhere (CO-RE)\cite{core} addresses varying kernel version issues by using BPF Type Format (BTF)\cite{btf} data, allowing eBPF programs to adapt dynamically and run on any compatible kernel without source code modifications, but this is still limited.

Compatibility across architectures introduces additional complexities, such as the kprobe and uprobe eBPF program uses \textbf{pt\_regs}, a structure that varies between different architectures. Differences in data structure layouts, endianness, and pointer widths also pose challenges.

\subsection{Deployment challenges in Cloud Environments}

Orchestrating the lifecycle of eBPF programs in the cloud is complex, involving various states such as loading, attaching, detaching, and unloading eBPF programs.

One common strategy is tight integration, where monitoring or management tools are integrated within the control plane application. This approach, used by projects like Cilium\cite{cilium}, Pixie\cite{pixie}, Tetragon\cite{tetragon}, and Deepflow\cite{shen2023network}, allows seamless interaction with system internals for efficient observation and manipulation of low-level operations. However, it has drawbacks. Deploying smaller eBPF tools or probes using traditional containers can be resource-consuming. This method often requires extensive permissions, posing security risks due to its extensive privileges\cite{he2023cross}. More granular capabilities like CAP\_PERFMON and CAP\_BPF have been introduced, but they still lack namespace constraints. Managing multi-user environments can also lead to conflicts where one eBPF program overrides another, resulting in silent failures or unpredictable behavior.

An alternative strategy is using Remote Procedure Calls (RPCs) to communicate between the control plane application and a dedicated BPF daemon. Tools like bpfman\cite{bpfman}, Inspektor-Gadget\cite{inspektor}, and bpfd\cite{bpfd} use this approach. The BPF daemon manages the BPF lifecycle and permissions, decoupling the BPF functionality from the application. While this adds a layer of abstraction, it also introduces challenges. The addition of a critical component in production increases the risk of failures, making troubleshooting and debugging more difficult. Maintaining consistency during updates can be problematic; when new kernel features are introduced, both the kernel dependency and the daemon need updates, which can delay the adoption of new capabilities. This model also imposes an additional support burden, as loaders must be compatible with tight integration and daemon delegation scenarios, complicating the upgrade or downgrade process and potentially leading to compatibility issues. Furthermore, operating system distributions or cloud providers may introduce different daemons, leading to a fragmented ecosystem.

\subsection{Challenges with Versioning and Pluggability}

Current eBPF tight coupling deployments make it difficult to version the eBPF program independently from its userspace counterpart. For example, suppose users must customize eBPF programs to track new proprietary protocols or analyze encrypted traffic with a specific userspace library. In that case, they must recompile the eBPF program and the userspace library, release a new version, and redeploy. This process becomes even more complex when the developer of the observability agent and the end user are from different organizations, as the developer may not have access to the proprietary protocols. 

Additionally, the lack of a standardized method for packaging and distributing eBPF programs leads to inconsistencies and management difficulties. Each user or organization may develop their own approach, resulting in fragmented and incompatible deployments. This lack of versioning and pluggability affects the flexibility and adaptability of eBPF programs, making updates cumbersome and error-prone. Without a modular and versioned approach, users cannot easily roll back to previous versions in case of issues, increasing the risk associated with deploying updates.

\section{Design of \sys}\label{sec:design}

This section details the design of \sys, which can package the eBPF application, include userspace control-plane code into a Wasm module, and deploy it either as a standalone container and utilizing existing container tools to deploy eBPF programs, or as an embedded runtime, managing eBPF application plugins within host applications.

\subsection{Architecture Overview}\label{sec:architecture}

\begin{figure}[t]
\centering
\includegraphics[width=0.5\textwidth]{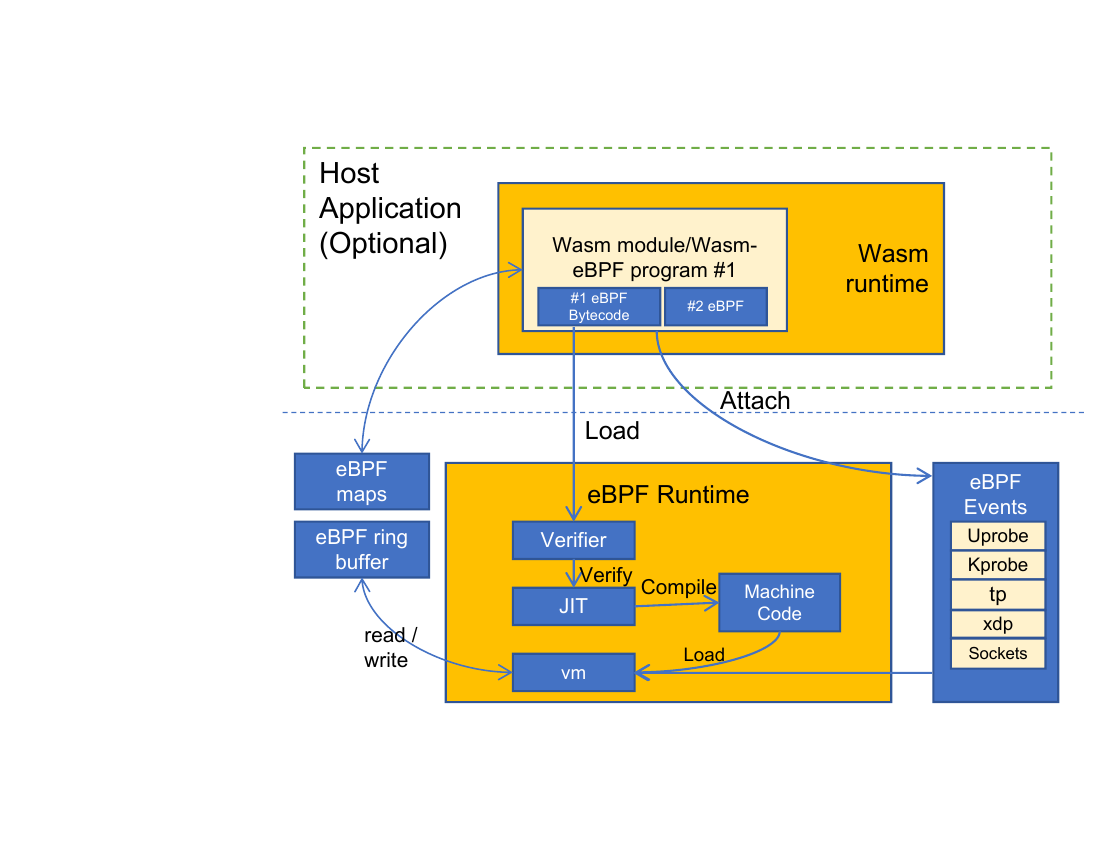}
\caption{The runtime of \sys}\label{fig:arch}
\end{figure}

The architecture of \sys addresses the challenges associated with deploying eBPF programs across diverse environments, managing their lifecycle in containers, and handling versioning and pluggability. \sys includes a WebAssembly (Wasm) library, toolchain, and runtime support for loading and executing eBPF programs, ensuring a consistent execution environment regardless of the underlying platform. \Cref{fig:arch} illustrates the runtime architecture of \sys.

In the runtime, a Wasm module can correspond to multiple eBPF programs. An eBPF program instance can be dynamically loaded into the kernel from the Wasm sandbox. This allows the selection of the desired attach point and control over the lifecycle of multiple eBPF bytecode objects. \sys supports various types of maps and enables bidirectional communication with the kernel, including support for most map types. It efficiently transfers information between kernel and user space through ring buffers polling (or vice versa) and map accesses. \sys is adaptable to nearly all use cases involving eBPF programs and can evolve and extend as kernel functionality evolves.

\subsection{\sys ABI Design}\label{sec:abi-design}

The ABI (Application Binary Interface) for \sys is designed to facilitate interaction between the Wasm runtime and eBPF programs. The ABI exports functions to the WebAssembly System Interface (WASI), the code snippet below illustrates the ABI functions provided by \sys.

\begin{figure}[t]
\centering
\includegraphics[width=0.5\textwidth]{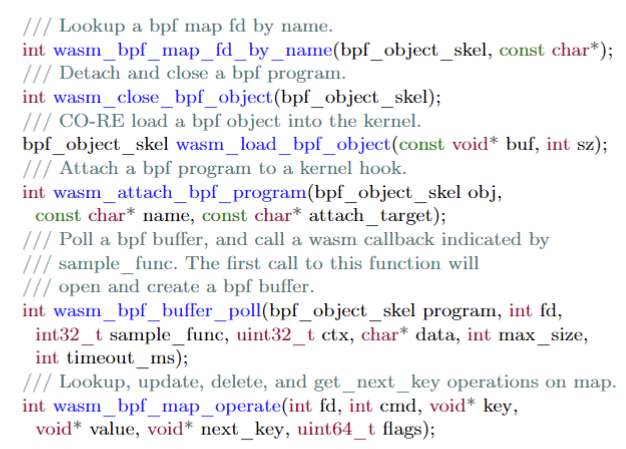}
\caption{\sys's ABI design}\label{fig:abi}
\end{figure}

We define the WASI ABI interface in \Cref{fig:abi} to ensure standardized interactions between eBPF programs and the Wasm runtime environment. The ABI functions cover operations such as loading eBPF objects, attaching eBPF programs to kernel hooks, polling eBPF buffers, and performing map operations. This design allows for robust and flexible management of eBPF programs within \sys, enabling their deployment and operation across various environments with minimal friction.

\subsection{Cross-Platform Compatibility}\label{sec:runtime-support-wasm}

While eBPF instructions themselves are inherently cross-platform, the userspace control applications that interact with the runtime are typically native binaries. This can lead to compatibility issues across different platforms. \sys addresses this challenge by leveraging WebAssembly (Wasm) and the WebAssembly System Interface (WASI) for userspace control plane applications, creating a robust and portable compatibility layer.

To ensure cross-platform access to function parameters in kprobe contexts, \sys incorporates a specific Compile Once – Run Everywhere (CO-RE) \cite{core} relocation pass for kprobe function parameters. This involves replacing pt\_regs data structures with newer architecture-specific versions and modifying the BTF\cite{btf} information of the program. Consequently, \sys can provide consistent access to function parameters in pt\_regs across different platforms.

\sys automatically selects the appropriate runtime—whether in the kernel or userspace—and eBPF features to handle the varying capabilities of different environments. This selection process ensures optimal performance and compatibility by adapting to the target environment's specific characteristics. For instance, in a Linux environment, \sys checks the kernel configuration and version to determine the available kernel features and then utilizes the suitable ones. Examples include using ring buffers to pass events to userspace in newer kernel versions and using perf events in older kernel versions. If kernel eBPF is not available and the eBPF program is intended for uprobe, \sys can automatically use bpftime\cite{zheng2023bpftime} as the userspace runtime.

\sys also enhances the CO-RE capabilities by incorporating mechanisms to ensure compatibility across different runtimes. This includes automatic BTF downloading in the runtime to ensure that necessary BTF data is available on the host, and provide BTF for userspace eBPF runtimes and host applications. As for a rich tapestry of kernel feature-interoperability issues, we require minor updates to the Wasm runtime code to make them compatible.

\sys integrates seamlessly with container toolchains\cite{docker-wasm} and Kubernetes. By replacing the standard containerd-shim with ctrd-wasmedge-shim and adhering to the specified implementation, \sys enables eBPF programs to be packaged as Open Container Initiative\cite{oci} (OCI) images and managed using standard container tools, streamlining the workflow for deploying and maintaining eBPF programs within cloud environments.



\subsection{Versioning and Pluggability}\label{sec:versioning-pluggability}

Leveraging Wasm toolchains, \sys packages eBPF applications into Wasm OCI images and uses the ORAS \cite{oras} tools to simplify storage and distribution in cloud environments. This standardized approach ensures consistent behavior across environments and simplifies large-scale management. The Wasm component model \cite{componentmodel} describes how wasm binary modules interact, allowing \sys applications to function as versioned plugins or libraries. This modularity facilitates updates and customizations without needing to recompile and redeploy the entire application. Independent versioning reduces deployment risks, enhances adaptability, and makes it easier to roll back to previous versions if issues arise.

\subsection{Performance Considerations}\label{sec:security-performance}

Ensuring minimal performance overhead while facilitating robust communication between Wasm and eBPF environments is a critical aspect of \sys. Given that Wasm can operate in both 32-bit and 64-bit modes, while eBPF is inherently 64-bit, differences in data structure layouts, endianness, and pointer widths can pose challenges. In the \sys project, communication between Wasm and eBPF virtual machines is optimized to eliminate the need for serialization and deserialization. Utilizing code generation techniques and support for BTF information in the toolchain, \sys ensures correct communication across different memory layouts with negligible runtime overhead. Data can be directly copied from kernel space to the Wasm virtual machine's memory, avoiding the additional overhead of multiple data transfers. Additionally, automatically generating eBPF program skeletons and type definitions enhances the development experience, making it more efficient and developer-friendly.

\section{Evaluation}
\label{sec:evaluation}

\begin{table*}[h]
\centering
\begin{tabular}{|l|c|c|c|c|c|}
\hline
\textbf{Program} & \textbf{Linux 5.5} & \textbf{Linux 6.10} & \textbf{Linux 6.10 arm64} & \textbf{Windows} & \textbf{Userspace eBPF} \\
\hline
bootstrap & -/O & X/O & X/O & -/- & -/- \\
lsm & -/- & X/O & X/O & -/- & -/- \\
opensnoop & X/O & X/O & X/O & -/- & -/O \\
sockops & X/O & X/O & X/O & -/O & -/- \\
kprobe & X/O & X/O & -/O & -/- & -/- \\
uprobe & X/O & X/O & -/O & -/- & -/O \\
xdp & X/O & X/O & X/O & -/- & -/O \\
\hline
\end{tabular}
\caption{Compatibility matrix showing support for various eBPF programs and features across different platforms for both native and \sys implementations.}
\label{tab:compatibility}
\end{table*}
We implemented \sys using approximately 1386 lines of Rust code and 510 lines of C++ code to answer the following research questions: the effectiveness of \sys and its compatibility. The toolchain includes a modified version of bpftool, which generates eBPF skeletons for development and facilitates serialization-free communication between the eBPF runtime and the Wasm runtime. To enable the development of eBPF programs within a Wasm environment, we implemented the \sys library in C, Go, and Rust (e.g., libbpf-wasm). This library interacts with the WASI interface and loads eBPF bytecode into the eBPF runtime. Additionally, we built the \sys runtime on top of WasmEdge\cite{long2020lightweight}, enabling interaction with the eBPF runtime and integration with container tools.

Our evaluation focuses on the following research questions:
\begin{itemize}
\item \textit{RQ1:} How effective is \sys at running control plane eBPF applications compared to native?
\item \textit{RQ2:} How does the start-up latency and binary size of \sys compare to standalone container deployments?
\item \textit{RQ3:} What is the compatibility of the eBPF programs across different platforms?
\end{itemize}

\subsection{Effectiveness}

The test environment includes an Intel Xeon E5-2697v2 @ 3.5GHz, with kernel version 6.6. We conducted micro-benchmarks to evaluate the performance of \sys compared to native eBPF implementations in \Cref{tab:micro-benchmarks}. \textbf{Map Access} measures the average time to access eBPF maps via syscall in both Wasm and native environments. \textbf{Ring Buffer Polling} measures the average latency per event for ring buffer polling operations in both Wasm and native environments. The results show that the Wasm abstraction introduces acceptable overhead compared to the native environment.

\begin{table}[h]
\centering
\begin{tabular}{|l|c|c|}
\hline
\textbf{Benchmark} & \textbf{Wasm (avg ns)} & \textbf{Native (avg ns)} \\
\hline
Map Access & 1885.26 & 1117.43 \\
Ring Buffer Polling & 3186.83 & 1509.18 \\
\hline
\end{tabular}
\caption{Micro benchmark results for map access and ring buffer polling operations comparing Wasm and native environments.}
\label{tab:micro-benchmarks}
\end{table}

\subsection{Container Deployment}

We measured the start-up latency using the bootstrap program, from the start of the container to the loading and attaching of eBPF programs. The results are presented in Table \ref{tab:start-up-latency}.

\begin{table}[hbtp]
\centering
\begin{tabular}{|c|c|}
\hline
\textbf{Wasm Lightweight Container} & \textbf{Docker} \\
\hline
 0.176 & 0.656 \\
\hline
\end{tabular}
\caption{Start-up latency for loading and attaching eBPF programs using the bootstrap program.}
\label{tab:start-up-latency}
\end{table}

We also compared the binary sizes of Docker and Wasm eBPF programs, considering the size of the minimal image for container deployments. The results are shown in Table \ref{tab:binary-sizes}. The results indicate that Wasm-based programs are significantly smaller compared to docker, making them more suitable for lightweight container deployments.

\begin{table}[hbtp]
\centering
\begin{tabular}{|l|r|r|}
\hline
\textbf{Program} & \textbf{Docker Size} & \textbf{Wasm Size} \\
\hline
bootstrap & 1.3M & 72K \\
execve & 1.3M & 37K \\
lsm & 1.3M & 46K \\
opensnoop & 1.3M & 64K \\
runqlat & 1.3M & 92K \\
sockfilter & 1.3M & 47K \\
sockops & 1.3M & 49K \\
uprobe & 1.3M & 45K \\
xdp & 1.3M & 44K \\
rust-bootstrap & 5.0M & 1.7M \\
tcpconnlat-libbpf-rs & 5.1M & 1.8M \\
\hline
\end{tabular}
\caption{Binary sizes of native and Wasm eBPF programs.}
\label{tab:binary-sizes}
\end{table}

\subsection{Compatibility}

To evaluate the compatibility of \sys, we tested various eBPF programs across different platforms, including multiple Linux versions, Windows, and userspace eBPF environments. The platforms considered include Linux 5.5, Linux 6.10, Linux 6.10 on ARM64 architecture, Windows, and userspace eBPF. The matrix in Table \ref{tab:compatibility} provides a view of the support for each eBPF program in both native and \sys environments. The symbols used in the table are "X" for native eBPF and "O" for \sys, while "-" indicates that the platform is not applicable.

The compatibility results demonstrate that \sys can effectively run a wide range of eBPF programs across different platforms, ensuring binary-level compatibility.

\section{Related Work}\label{sec:related}
WALI\cite{ramesh2023stop} extends the WebAssembly System Interface (WASI) by providing an interface for Wasm and the Linux kernel space to share resources. It supports various system calls, including \texttt{execv} and \texttt{fork}, along with auxiliary data. This extension enables a new class of virtualization where WebAssembly modules can interact with the host system more effectively. WebAssembly's control flow integrity guarantees provide an additional level of protection against remote code injection attacks for modules using WALI. Furthermore, capability-based APIs can be virtualized and implemented in terms of WALI, enhancing reuse and robustness through better layering. Our work efficiently extends the eBPF program to the WASI layer for getting all the toolchain benefits.

\section{Conclusion}\label{sec:conclusion}

This paper introduced \sys, a system leveraging WebAssembly (Wasm) and the WASI’s standardized system interface to address the challenges of deploying eBPF programs across diverse environments. \sys ensures cross-platform compatibility, eliminates serialization overhead and introduces minimal overhead. It integrates seamlessly with container tools, simplifying the deployment and orchestration of eBPF programs within containerized ecosystems.

\bibliographystyle{abbrvnat}

\bibliography{cite}

\end{document}